\documentclass[seceq]{ptptex}
\notypesetlogo                       
\title{
Ward Identities in the Derivation of Hawking Radiation from Anomalies%
}

\author{
Koichiro \textsc{Umetsu}$^{1,}$\footnote{E-mail: umetsu@phys.cst.nihon-u.ac.jp}%
}

\inst{
$^1$Institute of Quantum Science, College of 
Science and Technology,\\ Nihon University, Tokyo 101-8308, Japan}

\recdate{February 1, 2008}

\abst{
Robinson and Wilczek suggested a new method of deriving Hawking radiation 
by the consideration of anomalies.
The basic idea of their approach is that the flux of Hawking radiation is determined by 
anomaly cancellation conditions
in the Schwarzschild black hole (BH) background.
Iso et al. extended the method to
a charged Reissner-Nordstr\"om BH and a rotating Kerr BH, 
and they showed that the flux of Hawking radiation can also be determined 
by anomaly cancellation conditions 
and regularity conditions of currents at the horizon.
Their formulation gives the correct Hawking flux for all the cases at infinity 
and thus provides a new attractive method of understanding Hawking radiation.
We present some arguments clarifying for this derivation.
We show that the Ward identities and boundary conditions for covariant currents
without referring to the Wess-Zumino terms and the effective action are sufficient to 
derive Hawking radiation.
Our method, which does not use step functions, 
thus simplifies some of the technical aspects of the original formulation.
}

\begin{document}

\maketitle

\section{Introduction}
Hawking radiation is derived by using the quantum effect in black hole (BH) physics.
There are several methods of deriving of Hawking radiation.\cite{haw1,fulling,haw2} \ 
Hawking's original derivation, which calculates the Bogoliubov coefficients 
between the in- and out-states for 
a body collapsing to form a BH, is very direct and physical.\cite{haw1} \ 
It is well-known that the Hawking flux agrees with the blackbody flux at the temperature 
$T=\kappa/2\pi$, where $\kappa$ is the surface gravity of a BH, 
if we ignore the backscattering of particles falling into the horizon, 
i.e., the gray body radiation.

Robinson and Wilczek demonstrated a new method of deriving of Hawking radiation.\cite{rob1} \ 
They derived Hawking radiation by the consideration of quantum anomalies.
Their derivation has an important advantage 
in localizing the source of Hawking radiation 
near the horizon where anomalies are visible.
Since both of two anomalies and Hawking radiation are typical quantum effects,
it is natural that Hawking radiation is related to the anomalies in their derivation.
Iso et al. improved the approach of Ref.~\citen{rob1} 
and extended the method to a charged Reissner-Nordstr\"om BH.\cite{iso1} \ 
This approach was also extended to a rotating Kerr BH 
and a charged and rotating Kerr-Newman BH
by Murata and Soda \cite{mura} \ and
by Iso et al.\cite{iso2}.

The approach of Iso et al. \cite{iso2} \ is very transparent and interesting.
However, there remain several points to be clarified.
First, Iso et al. start by using both the consistent and covariant currents.
However, they only impose boundary conditions on covariant currents.
As discussed in Ref.~\citen{iso1}, 
it is not clear why we should use covariant currents instead of consistent ones
to specify the boundary conditions at the horizon.
Banerjee and Kulkarni considered an approach using only covariant currents
without consistent currents.\cite{bane1} \ 
However, their approach heavily relies on the Wess-Zumino terms defined
by consistent currents.\cite{Wess} \ 
The Wess-Zumino terms are also used in the approach of Iso et al.
Therefore, Banerjee and Kulkarni's approach is not completely described by covariant currents only.

Second,
in Iso et al.'s approach the region outside the horizon must be divided into two regions
because the effective theories are different near and far from the horizon.
They thus used step functions to divide these two regions.
We think that the region near the horizon 
and the region far from the horizon are continuously related.
Nevertheless, if one uses step functions,
terms with delta functions appear that originate from the derivatives of step functions
when one considers the variation of the effective action.
They disregarded the terms without delta functions by claiming that these terms correspond to
the contributions of the ingoing mode.
This is the second issue that we wish to address here.
Banerjee and Kulkarni also considered an approach
without step functions.\cite{bane2} \ 
They obtained the Hawking flux by using the effective actions and two boundary conditions 
for covariant currents.
However, they assumed that the effective actions are 2-dimensional 
in both the region near the horizon and that far from the horizon.\cite{polya,Leut} \ 
As discussed in Iso et al.'s approach,
the original 4-dimensional theory is the 2-dimensional effective theory 
in the region near the horizon.
However, the effective theory should be 4-dimensional 
in the region far from the horizon.

In contrast with the above approaches,
we derive the Hawking flux
using only the Ward identities and two boundary conditions for the covariant currents.
We formally perform the path integral, 
and the N\"other currents are constructed by the variational principle.
Therefore, we can naturally treat the covariant currents.\cite{fuji1} \ 
We do not use the Wess-Zumino term, the effective action or step functions.
Therefore, we do not need to define consistent currents.
Although we use the two boundary conditions used
in Banerjee and Kulkarni's method,
we use the 4-dimensional
effective theory far from the horizon 
and the 2-dimensional theory near the horizon.
In this sense,
our method corresponds to Iso et al.'s method.
It is easier to understand the derivation of 
the Ward identities directly from the variation of matter fields 
than their derivation from the effective action
since we consider Hawking radiation as resulting from the effects of matter fields.

Our approach is essentially based on Iso et al.'s approach.
However, we simplify the derivation of Hawking radiation by clarifying the above issues.
We only use the Ward identities and two boundary conditions for covariant currents, 
and we do not use the Wess-Zumino terms, the effective action or step functions.

The content of the paper is as follows.
In \S 2, we show our simple derivation of Hawking radiation in a rotating Kerr BH background.
Section 3 is devoted to conclusions and discussions. 
In Appendix A we show how to derive the Hawking flux 
in a charged Reissner-Nordst\"om BH using our approach.
\section{Simple derivation}
In this section, by clarifying the arguments used in the derivation in Ref.~\citen{iso2}, \ 
we show that we can obtain the same result by a simplified method.
Since we consider a rotating Kerr BH, the external space is given by the Kerr metric
\begin{align}
ds^2=&\frac{\triangle -a^2 \sin^2\theta}{\Sigma^2}dt^2
+\frac{2a\sin^2\theta}{\Sigma^2}(r^2+a^2-\triangle)dtd\varphi \notag\\
&+\frac{a^2\triangle \sin^2\theta-(r^2+a^2)^2}{\Sigma^2}\sin^2\theta d\varphi^2
-\frac{\Sigma^2}{\triangle}dr^2-\Sigma^2d\theta^2,
\label{kerr1}
\end{align}
where $a\equiv J/M$, $\triangle \equiv r^2-2Mr+ a^2=(r-r_+)(r-r_-)$ 
and $\Sigma^2\equiv r^2+a^2\cos^2 \theta$. $r_{+(-)}$ is the outer (inner) horizon.
We consider quantum fields in the vicinity of the Kerr BH.
In 4 dimensions, the action for a scalar field is given by
\begin{align}
S=\frac{1}{2}\int d^4 x\sqrt{-g}g^{\mu\nu}\partial_\mu\phi\partial_\nu \phi+S_{\rm int},
\label{kerract1}
\end{align}
where the first term is the kinetic term 
and the second term $S_{\rm int}$ represents the mass, 
potential and interaction terms.
Note that the $U(1)$ gauge field does not appear in the Kerr BH background.
This is a crucial difference from the charged BH.
By performing the partial wave decomposition of $\phi$ in terms of the spherical harmonics
($\phi=\sum_{l,m}\phi_{lm}Y_{lm}$) and using the properties of metrics at the horizon,
the action $S$ near the horizon is written as \cite{iso2}
\begin{align}
S=-\frac{1}{2}\sum_{l,m}\int dt dr(r^2+a^2)\phi_{lm}^*
\left[ \frac{(r^2+a^2)}{\triangle}\left( \partial_t  +\frac{ima}{r^2+a^2}\right)^2
-\partial_r \frac{\triangle}{r^2+a^2} \partial_r\right]\phi_{lm},
\label{kerract9}
\end{align}
where we ignore $S_{\rm int}$ because the kinetic term dominates near the horizon
in high-energy theory. 
From this action we find that $\phi_{lm}$ can be considered as $(1+1)$-dimensional complex
scalar fields in the backgrounds of the dilaton $\Phi$, metric $g_{\mu\nu}$ 
and $U(1)$ gauge field $A_\mu$, which are defined by
\begin{align}
&\Phi=r^2+a^2,
\label{dil}\\
&g_{tt}=f(r),\ 
g_{rr}=-\frac{1}{f(r)},\ 
g_{rt}=0,
\label{2met}\\
&A_t=-\frac{a}{r^2+a^2}, \ A_r=0,
\label{2gauge}
\end{align}
where $f(r)\equiv \triangle/(r^2+a^2)$.
The $U(1)$ charge of the 2-dimensional field $\phi_{lm}$ is $m$.

From (\ref{kerract9}), we find that the effective theory is 
the $(1+1)$-dimensional theory near the horizon.
However, we cannot simply regard the effective theory far from the horizon 
as the $(1+1)$-dimensional theory.
We need to divide the region outside the horizon into two regions
because the effective theories are different near the horizon
and far from the horizon.
We define region $O$ as the region far from the horizon and region $H$ as the region near the horizon.
Note that the action in region $O$ is
$S_{(O)}[\phi,g^{\mu\nu}_{(4)}]$
and 
the action in region $H$ is $S_{(H)}[\phi,g^{\mu\nu}_{(2)},A_\mu,\Phi]$.

\vspace{0.6cm}
\unitlength 0.1in
\begin{picture}( 22.0500, 18.3000)(  7.9500,-20.2000)
%
\special{pn 20}%
\special{pa 2200 400}%
\special{pa 800 1800}%
\special{fp}%
\special{pa 800 1800}%
\special{pa 800 1800}%
\special{fp}%
\special{pa 800 1800}%
\special{pa 800 1800}%
\special{fp}%
\special{pa 800 1800}%
\special{pa 800 1800}%
\special{fp}%
\put(12.0000,-8.0000){\makebox(0,0){{\bf BH}}}%
\put(22.2000,-3.6000){\makebox(0,0)[lb]{${\cal H}$}}%
%
\special{pn 8}%
\special{pa 2400 600}%
\special{pa 1000 2000}%
\special{dt 0.045}%
\special{pa 1000 2000}%
\special{pa 1000 2000}%
\special{dt 0.045}%
\special{pa 1000 2000}%
\special{pa 1000 2000}%
\special{dt 0.045}%
\special{pa 1000 2000}%
\special{pa 1000 2000}%
\special{dt 0.045}%
%
\special{pn 8}%
\special{pa 2200 1600}%
\special{pa 2400 1400}%
\special{fp}%
\special{sh 1}%
\special{pa 2400 1400}%
\special{pa 2340 1434}%
\special{pa 2362 1438}%
\special{pa 2368 1462}%
\special{pa 2400 1400}%
\special{fp}%
\special{pa 2400 1400}%
\special{pa 2400 1400}%
\special{fp}%
\special{pa 2400 1400}%
\special{pa 2400 1400}%
\special{fp}%
%
\special{pn 8}%
\special{pa 2200 1600}%
\special{pa 2000 1400}%
\special{fp}%
\special{sh 1}%
\special{pa 2000 1400}%
\special{pa 2034 1462}%
\special{pa 2038 1438}%
\special{pa 2062 1434}%
\special{pa 2000 1400}%
\special{fp}%
\special{pa 2000 1400}%
\special{pa 2000 1400}%
\special{fp}%
\special{pa 2000 1400}%
\special{pa 2000 1400}%
\special{fp}%
%
\special{pn 8}%
\special{pa 1400 1400}%
\special{pa 1600 1200}%
\special{fp}%
\special{sh 1}%
\special{pa 1600 1200}%
\special{pa 1540 1234}%
\special{pa 1562 1238}%
\special{pa 1568 1262}%
\special{pa 1600 1200}%
\special{fp}%
\special{pa 1600 1200}%
\special{pa 1600 1200}%
\special{fp}%
\special{pa 1600 1200}%
\special{pa 1600 1200}%
\special{fp}%
%
\special{pn 8}%
\special{pa 1400 1400}%
\special{pa 1200 1200}%
\special{dt 0.045}%
\special{sh 1}%
\special{pa 1200 1200}%
\special{pa 1234 1262}%
\special{pa 1238 1238}%
\special{pa 1262 1234}%
\special{pa 1200 1200}%
\special{fp}%
\special{pa 1200 1200}%
\special{pa 1200 1200}%
\special{dt 0.045}%
\special{pa 1200 1200}%
\special{pa 1200 1200}%
\special{dt 0.045}%
\put(12.0000,-16.0000){\makebox(0,0){{\bf $H$}}}%
\put(22.0000,-17.6000){\makebox(0,0){{\bf $O$}}}%
\put(30.0000,-13.8000){\makebox(0,0)[lb]{${\cal H}$; Future horizon}}%
\put(30.0000,-15.8000){\makebox(0,0)[lb]{Region $H$; 2 dimensions and chiral}}%
\put(30.0000,-17.8000){\makebox(0,0)[lb]{Region $O$; 4 dimensions and anomaly-free}}%
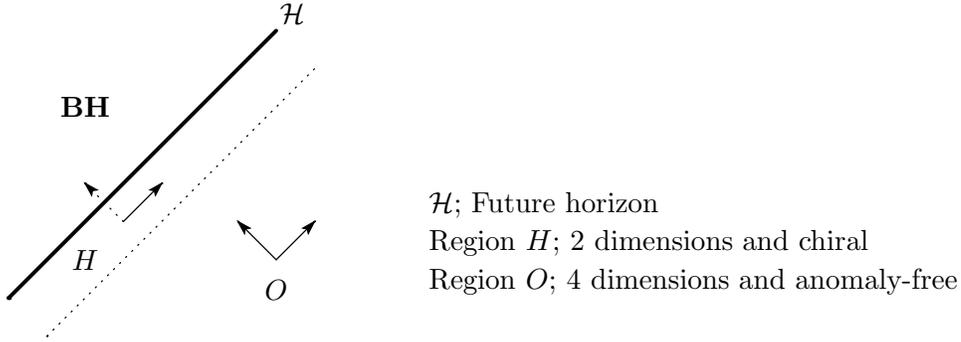
\put(12.1000,-21.9000){\makebox(0,0)[lb]{Fig. 1  Part of the Penrose diagram relevant to our analysis.}}%
\end{picture}%

\vspace{0.6cm}

The dashed arrow in region $H$ represents the ignored ingoing mode falling toward the horizon.
\vspace{0.6cm}

We can divide particles into ingoing modes falling toward the horizon (left-handed) 
and outgoing modes moving away from the horizon (right-handed) 
using a Penrose diagram\cite{rob1,iso1,iso2} (Fig. 1).
Since the horizon is a null hypersurface, none of the ingoing modes
at the horizon are expected to affect the classical physics outside the horizon.
Thus, we ignore the ingoing modes.
Therefore, anomalies appear with respect to the gauge or general coordinate symmetries 
since the effective theory is chiral near the horizon.
Here, we do not consider the backscattering of ingoing modes, i.e., the gray body radiation.

We now present the derivation of Hawking radiation for the Kerr BH.
First, we consider the effective theory in region $O$.
The effective theory is 4-dimensional in region $O$, which we cannot reduce to a 2-dimensional theory. 
In contrast with the case of a charged BH, 
a 4-dimensional gauge field such as the Coulomb potential $A=-Q/r$ 
does not exist in a rotating Kerr BH.
Therefore, we do not define the $U(1)$ gauge current in region $O$.
In contrast, the effective theory in region $H$ is a 2-dimensional chiral theory and
we can regard part of the metric as a gauge field such as (\ref{2gauge}),
since the action of (\ref{kerract9}) is $S_{(H)}[\phi,g^{\mu\nu}_{(2)},A_\mu,\Phi]$.

Second, we consider the Ward identity for the gauge transformation 
in region $H$ near the horizon.
Here, we pretend to formally perform the path integral for
$S_{(H)}[\phi,g^{\mu\nu}_{(2)},A_\mu,\Phi]$, 
where the N\"other current is constructed by the variational principle,
although we do not perform an actual path integral.
Therefore, we can naturally treat \textit{covariant} currents.\cite{fuji1} \ 
As a result, we obtain the Ward identity with a gauge anomaly
\begin{align}
\nabla_\mu J^{\mu}_{(H)}-{\cal C}=0,
\label{jcon1}
\end{align}
where we define covariant currents $J^\mu_{(H)}(r)$ 
and ${\cal C}$ is a covariant gauge anomaly.
This Ward identity is for right-handed fields.
The covariant form of the 2-dimensional abelian anomaly ${\cal C}$ is given by
\begin{align}
{\cal C}=\pm \frac{m^2}{4\pi\sqrt{-g_{(2)}}}\epsilon^{\mu\nu}F_{\mu\nu},\qquad(\mu,\nu=t,r)
\end{align}
where $+(-)$ corresponds to right(left)-handed matter fields, 
$\epsilon^{\mu\nu}$ is an antisymmetric tensor with $\epsilon^{tr}=1$ and
$F_{\mu\nu}$ is the field-strength tensor.
Using the 2-dimensional metric (\ref{2met}), (\ref{jcon1}) is written as
\begin{align}
\partial_r J^r_{(H)}(r)=\frac{m^2}{2\pi}\partial_r A_t(r).
\label{jcon22}
\end{align}
By integrating Eq. (\ref{jcon22}) over $r$ from $r_+$ to $r$, we obtain
\begin{align}
J^r_{(H)}(r)=\frac{m^2}{2\pi}\left[ A_t(r)- A_t(r_+)\right],
\label{cur2}
\end{align}
where we use the condition
\begin{align}
J^r_{(H)}(r_+)=0.
\label{reg1}
\end{align}
Condition (\ref{reg1}) corresponds to the statement
that free falling observers see a finite amount of the charged current at the horizon,
i.e., (\ref{reg1}) is derived from the regularity of covariant currents
(see the appendix of Ref.~\citen{iso2}).
We regard (\ref{cur2}) as a covariant $U(1)$ gauge current appearing
in region $H$ near the horizon.

Third, we consider the Ward identity for the general coordinate transformation 
in region $O$ far from the horizon.
By improving the approach of Ref.~\citen{iso1},
we define
the formal 2-dimensional energy-momentum tensor in region $O$ 
from the exact 4-dimensional energy-momentum tensor in region $O$
to connect the thus-defined 2-dimensional energy-momentum tensor in region $O$ 
with the 2-dimensional energy-momentum tensor in region $H$.
Since the action is $S_{(O)}[\phi,g^{\mu\nu}_{(4)}]$ in region $O$, 
the Ward identity for the general coordinate transformation is written as
\begin{align}
\nabla_\nu T^{\mu\nu}_{(4)}=0,
\label{con11}
\end{align}
where $T^{\mu\nu}_{(4)}$ is the 4-dimensional energy-momentum tensor.
Since the Kerr background is stationary and axisymmetric,
the expectation value of the energy-momentum tensor 
in the background depends only on $r$ and $\theta$,
i.e., $\langle T^{\mu\nu}\rangle=\langle T^{\mu\nu}(r,\theta) \rangle$.
The $\mu=t$ component of the conservation law (\ref{con11}) is written as
\begin{align}
\partial_r(\sqrt{-g} T^{~r}_{t(4)})+\partial_\theta (\sqrt{-g}T^{~\theta}_{t(4)})=0,
\label{4con1}
\end{align}
where $\sqrt{-g}=(r^2+a^2 \cos^2\theta)\sin \theta$.
By integrating Eq. (\ref{4con1}) over the angular coordinates $\theta$ and $\varphi$,
we obtain
\begin{align}
\partial_r T^{~r}_{t(2)}=0,
\end{align}
where we define the effective 2-dimensional tensor $T^{~r}_{t(2)}$ by
\begin{align}
T^{~r}_{t(2)}\equiv \int d\Omega_{(2)}(r^2+a^2\cos ^2\theta)T^{~r}_{t(4)}.
\label{trt2}
\end{align}
We define $T^{~r}_{t(2)}\equiv T^{~r}_{t(O)}$ to emphasize 
region $O$ far from the horizon.
The energy-momentum tensor $T^{~r}_{t(O)}$ is conserved in region $O$;
\begin{align}
\partial_r T^{~r}_{t(O)}=0.
\label{tcon1}
\end{align}
By integrating Eq. (\ref{tcon1}), we obtain
\begin{align}
T^{~r}_{t(O)}=a_o,
\label{tens1}
\end{align}
where $a_o$ is an integration constant.

Finally, we consider the Ward identity for the general coordinate transformation 
in region $H$ near the horizon.
The Ward identity for the general coordinate transformation when there is a gravitational anomaly
is 
\begin{align}
\nabla_\nu T^{~\nu}_{\mu(H)}(r)-F_{\mu\nu} J^\nu_{(H)}(r)
-\frac{\partial_\mu \Phi}{\sqrt{-g}}\frac{\delta S}{\delta \Phi}-{\cal A}_{\mu}(r)=0,
\label{tenward}
\end{align}
where both the gauge current and the energy-momentum tensor are defined
to be of the \textit{covariant} form
and ${\cal A}_\mu$ is the covariant form of the 2-dimensional gravitational anomaly.
This Ward identity corresponds to that of Ref.~\citen{bane1} when there is no dilaton field.
The covariant form of the 2-dimensional gravitational anomaly ${\cal A}_\mu$ 
is given by \cite{wit,fuji2,bert1}
\begin{align}
{\cal A}_\mu=\frac{1}{96\pi\sqrt{-g}}\epsilon_{\mu\nu}\partial^\nu R =\partial_r N^{~r}_{\mu},
\label{ano1}
\end{align}
where we define $N^{~r}_{\mu}$ by
\begin{align}
N^{~r}_{t}\equiv \frac{ff"-(f')^2/2}{96\pi},\qquad N^{~r}_{r}\equiv 0,
\label{nrt}
\end{align}
and \{ $'$ \} represents differentiation with respect to $r$.
The $\mu=t$ component of (\ref{tenward}) is written as
\begin{align}
\partial_r T^{~r}_{t(H)}(r)=F_{rt}J^r_{(H)}(r)+\partial_r N^{~r}_{t}(r).
\label{trth}
\end{align}
Using (\ref{cur2}) and integrating (\ref{trth}) over $r$ from $r_+$ to $r$, we obtain
\begin{align}
T^{~r}_{t(H)}(r)
=-\frac{m^2}{2\pi}A_t (r_+) A_t (r) + \frac{m^2}{4\pi} A^2_t(r)+N^{~r}_t(r)
+\frac{m^2}{4\pi} A^2_t(r_+)-N^{~r}_t(r_+),
\label{tens2}
\end{align}
where we impose the condition that the energy-momentum tensor vanishes at the horizon, 
which is the same as (\ref{reg1}):
\begin{align}
T^{~r}_{t(H)}(r_+)=0.
\label{reg2}
\end{align}

We compare (\ref{tens1}) with (\ref{tens2}).
By following Banerjee and Kulkarni's approach,\cite{bane2} \ 
we impose the condition that the asymptotic form of (\ref{tens2}) 
in the limit $r\rightarrow \infty$ is equal to (\ref{tens1}):
\begin{align}
T^{~r}_{t(O)}=T^{~r}_{t(H)}(\infty).
\label{condition2}
\end{align}
Condition (\ref{condition2}) corresponds to the statement that
no energy flux is generated away from the horizon region.
Therefore, the asymptotic form of (\ref{tens2}) has to agree with that of (\ref{tens1}).
From (\ref{condition2}), we can obtain
\begin{align}
a_o=\frac{m^2\Omega^2}{4\pi}+\frac{\pi}{12\beta^2},
\end{align}
where the quantity $\Omega$ is well known as the angular velocity in BH physics,\cite{bek}
\begin{align}
\Omega \equiv \frac{a}{r^2_+ + a^2},
\end{align}
and we use both the surface gravity of the BH,
\begin{align}
\kappa=\frac{2\pi}{\beta}=\frac{1}{2}f'(r_+),
\end{align}
and (\ref{nrt}).
As a result, we obtain the flux of the energy-momentum tensor 
in the region far from the horizon as
\begin{align}
T^{~r}_{t(O)}=\frac{m^2\Omega^2}{4\pi}+\frac{\pi}{12\beta^2}.
\label{sol1}
\end{align}
This flux agrees with the Hawking flux.
Our result corresponds to that of Ref.~\citen{iso2} in the limit $r\rightarrow\infty$.
In contrast with Ref.~\citen{iso2}, our result does not depend on gauge fields
in the region far from the horizon when the radial coordinate $r$ is large but finite.
As can be seen from the action (\ref{kerract1}),
the gauge field does not exist in the Kerr BH physics 
in a realistic 4-dimensional sense, and only the mass and angular momentum appear.
We thus consider that
our result presented here is more natural than that of Ref.~\citen{iso2}.
\section{Conclusion and discussion}
We have shown that the Ward identities and boundary conditions for covariant currents, 
without referring to the Wess-Zumino terms and the effective action, are sufficient to 
derive Hawking radiation.
The first boundary condition states 
that both the $U(1)$ gauge current and the energy-momentum tensor 
vanish at the horizon, as in (\ref{reg1}) and (\ref{reg2}).
This condition corresponds to the regularity condition
that a free falling observer sees a finite amount of the charged current at the horizon.
The second boundary condition is that
the asymptotic form of the energy-momentum tensor defined in the region near the horizon
is equal to the energy-momentum tensor 
in the region far from the horizon in the limit $r\rightarrow \infty$, 
as in (\ref{condition2}).
This condition means that no energy flux is generated far from the horizon.
In contrast with previous works,
we do not use the consistent current at any stage of our analysis
since we use neither the Wess-Zumino term nor the effective action.
We also do not use any step function.
Therefore, we believe that our approach clarifies some essential aspects of the derivation
of Hawking flux from anomalies.

When one compares our method with that of Iso et al. \cite{iso2}, \ 
one recognizes the following difference.
They defined the gauge current by the $\varphi$ component of 
the 4-dimensional energy-momentum tensor $T^{~r}_{\varphi(4)}$ 
in the region far from the horizon.
In contrast, we do not define the gauge current in the region far from the horizon, 
since no gauge current exists in a Kerr BH.
This difference appears in (\ref{tcon1}), whereas Iso et al. used the equation
\begin{align}
\partial_r T^{~r}_{t(2)}-F_{rt}J^r_{(2)}=0.
\label{miss1}
\end{align}
If we define gauge currents suitably,
we might be able to consider the Kerr BH 
in the same way as the Reissner-Nordstr\"om BH, as Iso et al. attempt to do.
However, some subtle aspects are involved in such methods
that attempt to define gauge currents.

To be explicit, the authors in Ref.~\citen{iso2} regard part of the metrics as the gauge field by defining 
$A^{\mu}\equiv -g^{\mu\varphi}_{(4)}$, as in Kaluza-Klein theory.
This definition is consistent with the initial definition of the current (\ref{2gauge})
near the horizon, i.e.,
\begin{align}
A_t=\frac{g_{t\varphi(4)}}{g_{\varphi\varphi(4)}}
=\frac{a(r^2+a^2 -\triangle)}{a^2 \triangle\sin^2 \theta -(r^2+a^2)^2}
\stackrel{\triangle\to 0}{\longrightarrow}
-\frac{a}{r^2+a^2}.
\end{align}
To maintain consistency, 
they simultaneously assume that 
the definition of (\ref{trt2}) is modified such that
it leads to (\ref{miss1}) by using 
the $\mu=t$ component of (\ref{con11}), i.e., 
\begin{align}
T_{t(2)}^{~r}=\int d \Omega_2 \left(r^2+a^2\cos^2 \theta\right) 
\left( T_{t(4)}^{~r} - A_t T^{~r}_{\varphi(4)} \right).
\end{align}
In this way they maintain consistency.
However, we consider that definition (\ref{trt2}) is more natural 
than this modified definition,
since in definition (\ref{trt2}) the formal 2-dimensional energy-momentum tensor 
is defined by
integrating the exact 4-dimensional energy-momentum tensor over the angular coordinates 
without introducing an artificial gauge current in the region far from the horizon. 
In our approach, which is natural for the Kerr BH, no gauge field appears
in the region far from the horizon where the radial coordinate $r$ is large but finite,
in contrast to the formulation in Ref.~\citen{iso2}.
We thus believe that our formulation is more natural than the formulation in Ref.~\citen{iso2}, 
although both formulations give rise to the same physical conclusion.

In passing, we mention that the Hawking flux is determined from (\ref{tens2}) simply by considering the direct limit
\begin{align}
T^{~r}_{t(H)}(r\rightarrow \infty)
=
\frac{m^2}{4\pi} A^2_t(r_+)-N^{~r}_t(r_+),
\end{align}
which agrees with (\ref{sol1}).
The physical meaning of this consideration is that Hawking radiation is induced by
quantum anomalies, which are defined in an arbitrarily small region near the horizon
since they are short-distance phenomena, 
and at any region far from the horizon the theory is anomaly-free
and thus, no further flux is generated.
Namely, we utilize an intuitive picture on the basis of the Gauss theorem, 
which is applied to a closed region surrounded 
by a surface $S$ very close to the horizon and a surface $S'$ far from the horizon
in the asymptotic region (Fig. 2).
If no flux is generated in this closed region,
the flux on the surface very close to the horizon
and the flux on the surface far from the horizon in the asymptotic region coincide.

\begin{center}
\unitlength 0.1in
\begin{picture}( 28.7100, 27.6700)( 11.2000,-30.2700)
%
\special{pn 8}%
\special{sh 0.600}%
\special{ar 2606 1648 360 360  0.0000000 6.2831853}%
%
\special{pn 8}%
\special{ar 2606 1648 468 468  0.0000000 0.0256410}%
\special{ar 2606 1648 468 468  0.1025641 0.1282051}%
\special{ar 2606 1648 468 468  0.2051282 0.2307692}%
\special{ar 2606 1648 468 468  0.3076923 0.3333333}%
\special{ar 2606 1648 468 468  0.4102564 0.4358974}%
\special{ar 2606 1648 468 468  0.5128205 0.5384615}%
\special{ar 2606 1648 468 468  0.6153846 0.6410256}%
\special{ar 2606 1648 468 468  0.7179487 0.7435897}%
\special{ar 2606 1648 468 468  0.8205128 0.8461538}%
\special{ar 2606 1648 468 468  0.9230769 0.9487179}%
\special{ar 2606 1648 468 468  1.0256410 1.0512821}%
\special{ar 2606 1648 468 468  1.1282051 1.1538462}%
\special{ar 2606 1648 468 468  1.2307692 1.2564103}%
\special{ar 2606 1648 468 468  1.3333333 1.3589744}%
\special{ar 2606 1648 468 468  1.4358974 1.4615385}%
\special{ar 2606 1648 468 468  1.5384615 1.5641026}%
\special{ar 2606 1648 468 468  1.6410256 1.6666667}%
\special{ar 2606 1648 468 468  1.7435897 1.7692308}%
\special{ar 2606 1648 468 468  1.8461538 1.8717949}%
\special{ar 2606 1648 468 468  1.9487179 1.9743590}%
\special{ar 2606 1648 468 468  2.0512821 2.0769231}%
\special{ar 2606 1648 468 468  2.1538462 2.1794872}%
\special{ar 2606 1648 468 468  2.2564103 2.2820513}%
\special{ar 2606 1648 468 468  2.3589744 2.3846154}%
\special{ar 2606 1648 468 468  2.4615385 2.4871795}%
\special{ar 2606 1648 468 468  2.5641026 2.5897436}%
\special{ar 2606 1648 468 468  2.6666667 2.6923077}%
\special{ar 2606 1648 468 468  2.7692308 2.7948718}%
\special{ar 2606 1648 468 468  2.8717949 2.8974359}%
\special{ar 2606 1648 468 468  2.9743590 3.0000000}%
\special{ar 2606 1648 468 468  3.0769231 3.1025641}%
\special{ar 2606 1648 468 468  3.1794872 3.2051282}%
\special{ar 2606 1648 468 468  3.2820513 3.3076923}%
\special{ar 2606 1648 468 468  3.3846154 3.4102564}%
\special{ar 2606 1648 468 468  3.4871795 3.5128205}%
\special{ar 2606 1648 468 468  3.5897436 3.6153846}%
\special{ar 2606 1648 468 468  3.6923077 3.7179487}%
\special{ar 2606 1648 468 468  3.7948718 3.8205128}%
\special{ar 2606 1648 468 468  3.8974359 3.9230769}%
\special{ar 2606 1648 468 468  4.0000000 4.0256410}%
\special{ar 2606 1648 468 468  4.1025641 4.1282051}%
\special{ar 2606 1648 468 468  4.2051282 4.2307692}%
\special{ar 2606 1648 468 468  4.3076923 4.3333333}%
\special{ar 2606 1648 468 468  4.4102564 4.4358974}%
\special{ar 2606 1648 468 468  4.5128205 4.5384615}%
\special{ar 2606 1648 468 468  4.6153846 4.6410256}%
\special{ar 2606 1648 468 468  4.7179487 4.7435897}%
\special{ar 2606 1648 468 468  4.8205128 4.8461538}%
\special{ar 2606 1648 468 468  4.9230769 4.9487179}%
\special{ar 2606 1648 468 468  5.0256410 5.0512821}%
\special{ar 2606 1648 468 468  5.1282051 5.1538462}%
\special{ar 2606 1648 468 468  5.2307692 5.2564103}%
\special{ar 2606 1648 468 468  5.3333333 5.3589744}%
\special{ar 2606 1648 468 468  5.4358974 5.4615385}%
\special{ar 2606 1648 468 468  5.5384615 5.5641026}%
\special{ar 2606 1648 468 468  5.6410256 5.6666667}%
\special{ar 2606 1648 468 468  5.7435897 5.7692308}%
\special{ar 2606 1648 468 468  5.8461538 5.8717949}%
\special{ar 2606 1648 468 468  5.9487179 5.9743590}%
\special{ar 2606 1648 468 468  6.0512821 6.0769231}%
\special{ar 2606 1648 468 468  6.1538462 6.1794872}%
\special{ar 2606 1648 468 468  6.2564103 6.2820513}%
\put(14.8000,-21.9000){\makebox(0,0)[rt]{$S'$}}%
%
\special{pn 8}%
\special{pa 2352 1898}%
\special{pa 1632 2618}%
\special{fp}%
\special{sh 1}%
\special{pa 1632 2618}%
\special{pa 1692 2586}%
\special{pa 1670 2580}%
\special{pa 1664 2558}%
\special{pa 1632 2618}%
\special{fp}%
%
\special{pn 8}%
\special{pa 2596 2008}%
\special{pa 2596 3028}%
\special{fp}%
\special{sh 1}%
\special{pa 2596 3028}%
\special{pa 2616 2960}%
\special{pa 2596 2974}%
\special{pa 2576 2960}%
\special{pa 2596 3028}%
\special{fp}%
%
\special{pn 8}%
\special{pa 2596 1280}%
\special{pa 2596 260}%
\special{fp}%
\special{sh 1}%
\special{pa 2596 260}%
\special{pa 2576 328}%
\special{pa 2596 314}%
\special{pa 2616 328}%
\special{pa 2596 260}%
\special{fp}%
%
\special{pn 8}%
\special{pa 2858 1396}%
\special{pa 3578 676}%
\special{fp}%
\special{sh 1}%
\special{pa 3578 676}%
\special{pa 3516 710}%
\special{pa 3540 714}%
\special{pa 3544 738}%
\special{pa 3578 676}%
\special{fp}%
%
\special{pn 8}%
\special{pa 2858 1900}%
\special{pa 3578 2620}%
\special{fp}%
\special{sh 1}%
\special{pa 3578 2620}%
\special{pa 3544 2560}%
\special{pa 3540 2582}%
\special{pa 3516 2588}%
\special{pa 3578 2620}%
\special{fp}%
%
\special{pn 8}%
\special{pa 2972 1638}%
\special{pa 3992 1640}%
\special{fp}%
\special{sh 1}%
\special{pa 3992 1640}%
\special{pa 3924 1620}%
\special{pa 3938 1640}%
\special{pa 3924 1660}%
\special{pa 3992 1640}%
\special{fp}%
%
\special{pn 8}%
\special{pa 2252 1638}%
\special{pa 1234 1640}%
\special{fp}%
\special{sh 1}%
\special{pa 1234 1640}%
\special{pa 1302 1660}%
\special{pa 1288 1640}%
\special{pa 1302 1620}%
\special{pa 1234 1640}%
\special{fp}%
%
\special{pn 8}%
\special{pa 2352 1412}%
\special{pa 1632 692}%
\special{fp}%
\special{sh 1}%
\special{pa 1632 692}%
\special{pa 1664 754}%
\special{pa 1670 730}%
\special{pa 1692 726}%
\special{pa 1632 692}%
\special{fp}%
\put(29.2000,-24.7000){\makebox(0,0){$O$}}%
\put(26.0000,-16.4000){\makebox(0,0){\textbf{BH}}}%
%
\special{pn 8}%
\special{ar 2600 1640 1220 1220  0.0000000 6.2831853}%
%
\special{pn 8}%
\special{ar 2600 1650 500 500  0.0000000 6.2831853}%
\put(28.0000,-20.2000){\makebox(0,0){$H$}}%
\put(21.0000,-18.4000){\makebox(0,0)[rt]{$S$}}%
\end{picture}%

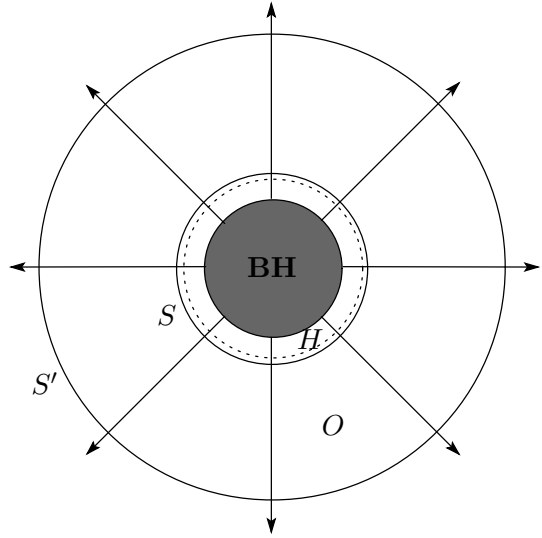
Fig. 2 Intuitive picture on the basis of the Gauss theorem.
\end{center}
The flux is only generated inside the dashed line.
The total fluxes on $S$ and $S'$ are equal from the Gauss theorem.
\vspace{0.6cm}

Finally, we discuss why we use the regularity conditions 
for \textit{covariant} currents instead of consistent currents.
All the physical quantities should be gauge-invariant.
Thus, physical currents should be \textit{covariant}.
This is consistent with, for example, the well-known anomalous baryon number current 
in the Weinberg-Salam theory.\cite{t'hooft}

For recent related works, please refer to Refs.~\citen{iso3, ind, chi}.

\section*{Acknowledgements}
I would like to thank my supervisor, K. Fujikawa, for helpful comments
and for carefully reading the present manuscript.
I also wish to acknowledge valuable discussions with 
S. Iso, H. Umetsu, R. Banerjee 
and the participants of the RIKEN theory workshop, December 22--23, 2007. 
%
\appendix
\section{The Case for a Charged BH} 
In this appendix,
we show that Hawking flux can be obtained in a charged BH using our approach.
Since we consider a charged Reissner-Nordstr\"om BH,
the external space is given by the Reissner-Nordstr\"om metric
\begin{align}
ds^2=f(r)dt^2-\frac{1}{f(r)}dr^2-r^2 d\theta^2-r^2\sin^2\theta d\varphi^2,
\end{align}
and $f(r)$ is given by
\begin{align}
f(r)=1-\frac{2M}{r}+\frac{Q^2}{r^2}=\frac{(r-r_+)(r-r_-)}{r^2},
\end{align}
where $r_\pm=M\pm \sqrt{M^2-Q^2}$
and $r_+$ is the distance from the center of the BH to the outer horizon.
We consider quantum fields in the vicinity of the Reissner-Nordstr\"om BH.
In 4 dimensions, the action for a complex scalar field is given by
\begin{align}
S=\int d^4x\sqrt{-g}g^{\mu\nu}(\partial_\mu+ieA_\mu)\phi^*(\partial_\nu-ieA_\nu)\phi
+S_{\rm int},
\label{srei1}
\end{align}
where the first term is the kinetic term 
and the second term $S_{\rm int}$ represents the mass, potential and interaction terms.
In contrast with the Kerr BH background,
note that the $U(1)$ gauge field $A_t=-Q/r$ appears
in the Reissner-Nordstr\"om BH background.
By performing the partial wave decomposition of $\phi$ in terms of the spherical harmonics
($\phi=\sum_{l,m}\phi_{lm}Y_{lm}$) and using the property $f(r_+)=0$ at the horizon,
the action $S$ near the horizon is written as
\begin{align}
S=-\sum_{l,m}\int dtdr r^2\phi^*_{lm}\left[\frac{1}{f(r)}(\partial_t-ieA_t)^2
-\partial_r f(r) \partial_r \right]\phi_{lm},
\end{align}
where we ignore $S_{\rm int}$ 
because the kinetic term dominates near the horizon in high-energy theory. 
From this action we find that $\phi_{lm}$ can be considered as $(1+1)$-dimensional complex
scalar fields in the backgrounds of the dilaton $\Phi$, metric $g_{\mu\nu}$ 
and $U(1)$ gauge field $A_\mu$, where
\begin{align}
&\Phi=r^2,\\
&g_{tt}=f(r),\quad
g_{rr}=-\frac{1}{f(r)},\quad
g_{rt}=0,
\label{grr2}\\
&A_t=-\frac{Q}{r}, \ A_r=0.
\end{align}
The $U(1)$ charge of the 2-dimensional field $\phi_{lm}$ is $e$.
Note that the action in the region far from the horizon 
is $S_{(O)}[\phi,g^{\mu\nu}_{(4)},A_\mu]$ and 
the action in the region near the horizon is $S_{(H)}[\phi,g^{\mu\nu}_{(2)},A_\mu,\Phi]$.

We now present the derivation of Hawking radiation for the Reissner-Nordstr\"om BH.
First, we consider the Ward identity for the gauge transformation
in region $O$ far away from the horizon.
Here, we formally perform the path integral for $S_{(O)}[\phi,g^{\mu\nu}_{(4)},A_\mu]$,
where the N\"other current is constructed by the variational principle.
Therefore, we can naturally treat \textit{covariant} currents.\cite{fuji1} \ 
As a result, we obtain the Ward identity
\begin{align}
\nabla_\mu J^\mu_{(4)}=0,
\label{Jcon1}
\end{align}
where $J^\mu_{(4)}$ is the 4-dimensional gauge current.
Since the Reissner-Nordstr\"om background is stationary and spherically symmetric,
the expectation value of the gauge current in the background depends only on $r$,
i.e., $\langle J^{\mu} \rangle=\langle J^{\mu}(r) \rangle$.
Using the 4-dimensional metric, the conservation law (\ref{Jcon1}) is written as
\begin{align}
\partial_r(\sqrt{-g}J^r_{(4)})+(\partial_\theta\sqrt{-g})J^\theta_{(4)}=0,
\label{Jcon2}
\end{align}
where $\sqrt{-g}=r^2 \sin \theta$.
By integrating Eq. (\ref{Jcon2}) over the angular coordinates $\theta$ and $\varphi$, 
we obtain
\begin{align}
\partial_r J^r_{(2)}=0,
\end{align}
where we define the effective 2-dimensional current $J^r_{(2)}$ by
\begin{align}
J^r_{(2)}\equiv\int d\Omega_{(2)} r^2 J^r_{(4)}.
\label{2.21}
\end{align}
We define $J^r_{(2)}\equiv J^r_{(O)}$ to emphasize region $O$ 
far from the horizon.
The gauge current $J^r_{(O)}$ is conserved in region $O$,
\begin{align}
\partial_r J^r_{(O)}=0.
\label{2.20}
\end{align}
By integrating Eq. (\ref{2.20}), we obtain
\begin{align}
J^r_{(O)}=c_o,
\label{2.23}
\end{align}
where $c_o$ is an integration constant.

Second, we consider the Ward identity for gauge transformation 
in the region $H$ near the horizon.
The Ward identity for the gauge transformation when there is a gauge anomaly 
is given by
\begin{align}
\nabla_\mu J^\mu_{(H)}-{\cal B}=0,
\label{2.24}
\end{align}
where we define the covariant current as $J^\mu_{(H)}$
and ${\cal B}$ is a covariant gauge anomaly.
The covariant form of the 2-dimensional gauge anomaly ${\cal B}$ is given by
\begin{align}
{\cal B}=\pm \frac{e^2}{4\pi\sqrt{-g}}\epsilon^{\mu\nu}F_{\mu\nu},\qquad (\mu,\nu=t,r)
\end{align}
where $+(-)$ corresponds to the anomaly for right(left)-handed fields.
Here $\epsilon^{\mu\nu}$ is an antisymmetric tensor with $\epsilon^{tr}=1$ and
$F_{\mu\nu}$ is the field-strength tensor.
Using the 2-dimensional metric (\ref{grr2}), (\ref{2.24}) is written as
\begin{align}
\partial_r J^r_{(H)}(r)=\frac{e^2}{2\pi}\partial_r A_t(r).
\label{a16}
\end{align}
By integrating (\ref{a16}) over $r$ from $r_+$ to $r$, we obtain
\begin{align}
J^r_{(H)}(r)=\frac{e^2}{2\pi}[ A_t(r)- A_t(r_+)],
\label{2.29}
\end{align}
where we impose the condition
\begin{align}
J^r_{(H)}(r_+)=0.
\end{align}
This condition corresponds to (\ref{reg1}) in the present paper.
We also impose the condition that the asymptotic form of (\ref{2.29}) 
is equal to that of (\ref{2.23}),
\begin{align}
J^r_{(O)}(\infty)=J^r_{(H)}(\infty).
\label{a19}
\end{align}
From (\ref{a19}), we obtain the gauge current in region $O$ as
\begin{align}
J^r_{(O)}=-\frac{e^2}{2\pi}A_t(r_+).
\label{2.31}
\end{align}

Third, we consider the Ward identity for the general coordinate transformation 
in the region $O$ far from the horizon.
We define the formal 2-dimensional energy-momentum tensor in region $O$ 
from the exact 4-dimensional energy-momentum tensor in region $O$
to connect the thus-defined 2-dimensional energy-momentum tensor in region $O$ 
with the 2-dimensional energy-momentum tensor in region $H$.
Since the action is $S_{(O)}[\phi,g^{\mu\nu}_{(4)},A_\mu]$ in region $O$, 
the Ward identity for the general coordinate transformation is written as
\begin{align}
\nabla_\nu T^{~\nu}_{\mu(4)}-F_{\nu\mu}J^\nu_{(4)}=0,
\label{a21}
\end{align}
where $T^{\mu\nu}_{(4)}$ is the 4-dimensional energy-momentum tensor.
Since the Reissner-Nordstr\"om background is stationary and spherically symmetric,
the expectation value of the energy-momentum tensor in the background depends only on $r$,
i.e., $\langle T^{\mu\nu}\rangle=\langle T^{\mu\nu}(r) \rangle$.
The $\mu=t$ component of the conservation law (\ref{a21}) is written as
\begin{align}
\partial_r \left( \sqrt{-g} T^{~r}_{t(4)}\right)
+\left( \partial_\theta \sqrt{-g} \right) T^{~\theta}_{t(4)}
-\sqrt{-g}F_{r t}J^r_{(4)}=0.
\label{a22}
\end{align}
By integrating (\ref{a22}) over $\theta$ and $\varphi$, we obtain
\begin{align}
\partial_r T^{~r}_{t(2)}=F_{r t}J^r_{(2)},
\label{2.34}
\end{align}
where we define the effective 2-dimensional tensor $T^{~r}_{t(2)}$ by
\begin{align}
T^{~r}_{t(2)}\equiv \int d\Omega_{(2)} r^2 T^{~r}_{t(4)},
\end{align}
and $J^r_{(2)}$ is defined by (\ref{2.21}).
To emphasize region $O$ far from the horizon,
we write (\ref{2.34}) as
\begin{align}
\partial_r T^{~r}_{t(O)}=F_{r t}J^r_{(O)}.
\label{a25}
\end{align}
By substituting (\ref{2.31}) into (\ref{a25}) and integrating it over $r$, we obtain
\begin{align}
T^{~r}_{t(O)}(r)=a_o-\frac{e^2}{2\pi}A_t(r_+)A_t(r).
\label{2.39}
\end{align}

Finally, we consider the Ward identity for the general coordinate transformation 
in region $H$ near the horizon.
The Ward identity for the general coordinate transformation when there is a gravitational anomaly
is 
\begin{align}
\nabla_\nu T^{~\nu}_{\mu(H)}-F_{\nu\mu}J^\nu_{(H)}-
\frac{\partial_\mu \Phi}{\sqrt{-g}}\frac{\delta S}{\delta \Phi}-{\cal A}_\mu=0,
\label{2.40}
\end{align}
where both the gauge current and the energy-momentum tensor 
are defined to be of the \textit{covariant} form
and ${\cal A}_\mu$ is the covariant form of the 2-dimensional gravitational anomaly.
This Ward identity corresponds to that of Ref.~\citen{bane1} when there is no dilaton field.
The covariant form of the 2-dimensional gravitational anomaly ${\cal A}_\mu$ 
agrees with (\ref{ano1}).
Using the 2-dimensional metric (\ref{grr2}),
the $\mu=t$ component of (\ref{2.40}) is written as
\begin{align}
\partial_r T^{~r}_{t(H)}(r)=\partial_r\left[-\frac{e^2}{2\pi}A_t(r_+)A_t(r)
+\frac{e^2}{4\pi}A^2_t(r)+N^{~r}_{t}\right].
\label{a28}
\end{align}
By integrating (\ref{a28}) over $r$ from $r_+$ to $r$, we obtain
\begin{align}
T^{~r}_{t(H)}(r)=-\frac{e^2}{2\pi}A_t(r_+) A_t(r)+\frac{e^2}{4\pi}A_t^2(r)+N^{~r}_{t}(r)
+\frac{e^2}{4\pi}A^2_t(r_+)-N^{~r}_t(r_+),
\label{a29}
\end{align}
where we impose the condition that the energy-momentum tensor vanishes at the horizon, 
which is the same as (\ref{reg2}):
\begin{align}
T^{~r}_{t(H)}(r_+)=0.
\end{align}
As for (\ref{condition2}),
we impose the condition that the asymptotic form of (\ref{a29}) 
in the limit $r\rightarrow \infty$ is equal to that of (\ref{2.39}),
\begin{align}
T^{~r}_{t(O)}(\infty)=T^{~r}_{t(H)}(\infty).
\label{a31}
\end{align}
From (\ref{a31}), we obtain
\begin{align}
a_o=\frac{e^2}{4\pi}A^2_t(r_+)-N^{~r}_t(r_+).
\end{align}
We thus obtain the flux of the energy-momentum tensor 
in the region far from the horizon as 
\begin{align}
T^{~r}_{t(O)}(r)=\frac{e^2 Q^2}{4\pi r^2_+}+\frac{\pi}{12\beta^2}+\frac{e^2Q}{2\pi r_+}A_t(r).
\end{align}
This result agrees with that of Ref.~\citen{iso1}\footnote{For the generalization of 
the present analysis to higher-spin currents, 
see Ref.~\citen{iso3}.}.
In contrast with the case of a rotating Kerr BH,
the energy flux depends on the gauge field 
in the region far from the horizon, 
but the radial coordinate $r$ is still finite
since the gauge field exists in a charged Reissner-Nordstr\"om BH background.
However, in the evaluation of Hawking radiation by setting $r\rightarrow \infty$,
the effect of the gauge field vanishes.

%

\end{document}